\documentclass[useAMS,usenatbib]{mn2e}
\usepackage{array}
\usepackage{amsmath}
\usepackage{graphicx}
\begin{document}

\title[Magnification in SDSS]{Magnification of Photometric LRGs by Foreground LRGs and Clusters in SDSS}

\author[Bauer et al.]{Anne H. Bauer$^{1}$\thanks{E-mail:
bauer@ieec.uab.es}, Enrique Gazta\~naga$^{1}$, Pol Mart{\'{\i}}$^{2}$, Ramon Miquel$^{2,3}$ \\ 
$^1$Institut de Ci\`encies de l'Espai, CSIC/IEEC, E-08193 Bellaterra, Spain \\
$^2$Institut de F\'isica d'Altes Energies, Universitat Aut\`onoma de Barcelona, E-08193 Bellaterra, Spain \\
$^3$Instituci\'o Catalana de Recerca i Estudis Avan\c{c}ats, E-08010 Barcelona, Spain}
\pagerange{\pageref{firstpage}--\pageref{lastpage}} \pubyear{2013}

\maketitle

\label{firstpage}

\begin{abstract}

The magnification effect of gravitational lensing is a powerful probe of the distribution of matter in the universe, yet it is frequently 
overlooked due to the fact that its signal to noise is smaller than that of lensing shear.  Because its systematic errors are 
quite different from those of shear, magnification is nevertheless an important approach with which to study the 
distribution of large scale structure.  We present lensing mass profiles of spectroscopic luminous red galaxies (LRGs) and 
galaxy clusters determined through measurements of the weak lensing magnification of 
photometric LRGs in their background.  We measure the change in detected galaxy counts as well as the increased average galaxy flux behind the lenses.  
In addition, we examine the average change in source color due to extinction by dust in the lenses.  By simultaneously fitting these three probes 
we constrain the mass profiles and dust-to-mass ratios of the lenses in six bins of lens richness.  For each richness bin we fit an NFW halo mass, 
brightest cluster galaxy (BCG) mass, second halo term, 
and dust-to-mass ratio.  The resulting mass-richness 
relation is consistent with previous analyses of the catalogs, and limits on the dust-to-mass ratio in the lenses are in agreement 
with expectations.  We explore the effects of including the (low signal-to-noise) flux magnification and reddening measurements in the analysis 
compared to using only the counts magnification data; the additional probes significantly improve the agreement between our measured 
mass-richness relation and previous results.

\end{abstract}


\section{Introduction}

Gravitational lensing, the deflection of photons by changes in gravitational potential, modifies the observed brightness and shape of 
sources behind massive objects.  Lensing is therefore a powerful technique to measure the distribution of matter, as its effects on source flux 
and shape are directly due to the mass along the line of sight.  

Galaxy clusters are extremely massive collapsed systems and act as powerful gravitational lenses.  As the peaks of the matter density distribution, 
they constitute sensitive probes of cosmology \citep[e.g.,][]{mantz08, rozo10}.  To constrain cosmology using the abundance of galaxy clusters, 
one must have accurate knowledge of the 
clusters' masses; gravitational lensing provides a natural way to constrain these masses.

Traditionally, weak gravitational lensing has been measured using shear: the distortion of the shape of galaxies behind a lens.  Shear is 
technically very difficult to measure, requiring high-quality, resolved images of distant galaxies;  as a result, increasing attention is
being paid to the magnification effects of weak lensing.  The signal-to-noise of the magnification of source galaxies behind a lens 
is typically smaller than that of the galaxies' shear \citep[see][equ. 7]{schneider00}.  The exact ratio depends on the shape of the lens mass 
profile and the sources' lensing sensitivity $\alpha_{c}$, defined below in equation \ref{alpha_c_eq}, but for a singular isothermal sphere lens 
and a given number of sources with the same $\alpha_{c}$ as in this work, the signal-to-noise of the magnification effect 
is about a factor of three smaller than that of lensing shear.  In magnification analyses, however, one can include unresolved (fainter, higher-redshift) galaxies in the source sample and thereby 
reduce the statistical noise in the result.  In addition, it is important to study magnification even if its statistical errors 
are larger than those of shear, as the systematic effects that complicate the measurement of 
magnification (e.g., photometric calibration homogeneity, detection completeness) are typically different from those afflicting shear 
(e.g., image resolution and seeing, intrinsic alignments).

Magnification is a fundamentally different lensing probe from shear in that its effects are proportional to the lens mass profile itself, 
as opposed to the change in the mass distribution as in the case of shear.  Magnification, in the weak lensing regime, measures the divergence of the 
gradient of the lensing deflection potential (the scalar convergence $\kappa$), while the shear is a polar given by the second derivatives of the potential.  
As a result, certain profile shapes (e.g. a point mass, or the 
profile of a second halo term) appear very differently to the two probes, allowing the two lensing signals to provide complementary information.  
In addition, shear 
measurements suffer from the ``mass-sheet degeneracy'', whereby a uniform sheet of mass is undetectable.  This degeneracy complicates the use of 
shear to calibrate the masses of lenses, as the point where the shear signal falls to zero may not be the point where the mass overdensity falls to 
zero.  The degeneracy can be practically overcome by using large areas of sky over which one may assume that cosmic variance does not 
cause a significant mass sheet.  Still, the problem is formally eliminated by using magnification measurements, where a mass sheet causes a 
measurable effect.  Magnification is therefore important in ensuring the proper normalization of mass profiles in lensing analyses.

Most studies of weak lensing magnification have measured ``magnification bias'': the change in detection rate of a magnified source population 
in a flux-limited survey due to the combination of increased observed object flux and decreased solid angle behind a lens.  Depending on how the number 
count distribution of the source sample changes with luminosity, magnification bias can lead to an increase or decrease in counts.  Other studies have 
measured the increase in measured flux \citep{menard09} and size \citep{schmidt12} in lensed sources, and how these lensed quantities affect scaling relations 
such as the fundamental plane \citep{sonnenfeld11, huff11} and the quasar variability-luminosity relation \citep{agn_magn, bauer12}.  
Studying another consequence of lensing magnification, \cite{coupon13} measured the change in the distribution of spectroscopic redshifts measured behind lenses.

When measuring lensing magnification it is essential to minimize redshift overlap in the source and lens samples, as physical correlations 
between the source and lens objects would lead to an increase of objects in the source catalog as a function of decreasing angular distance from 
objects in the lens catalog.  Depending on the properties of the source catalog, magnification can induce a positive or negative change in source number count 
density (see Section \ref{counts_section}).  Redshift overlap between the catalogs can therefore mimic or counteract the true lensing signal, with 
an amplitude much larger than that of magnification.  Because of this, photometric source samples in magnification analyses 
have consisted of objects with very reliable estimates of high redshifts, such as quasars \citep{gaztanaga03, scranton05, menard09}, 
Lyman break galaxies \citep{ludo10, hildebrandt11, hildebrandt12, morrison12, ford12, ford13}, and galaxies in targeted fields with unusually deep, 
high-quality photometry \citep{schmidt12, umetsu12}.  In this work we use a sample of luminous red galaxies (LRGs) for which we have well-measured photometric redshifts.  
These galaxies are at a redshift lower than is typical in magnification analyses (0.5$<$z$<$0.6), but as we have good knowledge of the objects' true redshift 
distribution we can quantify any contamination with our lens samples.

Although gravitational lensing by a massive object increases a source's observed flux, absorption by dust associated with the same lens can do the opposite.  
Dust absorption also causes a reddening of the source flux.  Because lensing is achromatic, it is possible to disentangle the two effects and study both 
the lens mass and dust content using multi-color data \citep{menard09}.

We measure the weak lensing magnification of the photometric MegaZ sample of LRGs \citep{megazdr7} from the Sloan Digital Sky 
Survey Data Release 7 \citep{sdss7}, due to LRG \citep{cabre09} and galaxy cluster \citep{dr9sz} lenses in the foreground.  Our lens samples are 
spectroscopic, with redshifts from 0.1 to 0.4.  The sources have photometric redshifts between 0.5 and 0.6.  We simultaneously measure 
the counts increase and average flux increase due to lensing, as well as any reddening due to dust in the lenses.  Using these probes we constrain 
the mass profiles of the lenses as modeled as a dark matter halo, brightest cluster galaxy, and second halo term, and also put limits on the 
dust-to-mass ratio in the lenses.  We fit such lens models in each of six bins in lens richness: one bin for the LRG sample, and five for the cluster catalog.

Throughout the analyses we assume the cosmological parameters 
$H_{0} = 70 \mathrm{km}\  \mathrm{s}^{-1} \mathrm{Mpc}^{-1}, \Omega_{m} = 0.3, 
\Omega_{\Lambda} = 0.7, \sigma_{8} = 0.807, n_{s} = 0.961$.

\section{Lensing Magnification}

The deflection of light by a gravitational lens causes sources in the lens' background to appear brighter than they 
intrinsically are.  This is due to the apparent geometrical stretching of space behind the lens, coupled with the 
fact that lensing conserves the sources' apparent surface brightness.  The magnification $\mu$ can be calculated from the 
Jacobian matrix $\mathcal A$ of the lensing distortion:

\begin{equation}
	\mu = \frac{1}{\mathrm{det} \mathcal{A}} = \frac{1}{(1 - \kappa)^{2} - | \gamma | ^{2}}
\label{mudef_eq}
\end{equation}

where $\kappa$ is the convergence and $\gamma$ is the shear induced by the lens.  For a general description 
of gravitational lensing physics, see, e.g., \cite{schneider04}.  

In the limit of weak lensing, where $\kappa$ and $\gamma$ are small, equation \ref{mudef_eq} can be approximated by

\begin{equation}
	\mu \approx 1 + 2 \kappa.
\label{muapprox_eq}
\end{equation}

The dimensionless convergence $\kappa$ can be expressed as the surface mass density $\Sigma$ divided by 
the critical surface mass density of the lensing system $\Sigma_{cr}$:
\begin{equation}
\kappa \equiv \frac{\Sigma}{\Sigma_{cr}}
\label{kappa_eq}
\end{equation}

with

\begin{equation}
	\Sigma_{cr} = \frac{c^{2}}{4 \pi G} \frac{D_{s}}{D_{d} D_{ds}}
\end{equation}

where $D_{d}$, $D_{s}$, $D_{ds}$ are the angular diameter distances from the observer to the lens, 
the observer to the source, and from the lens to the source, respectively.

Because we typically deal with fluctuations in magnification around $\mu = 1$, it is convenient to define 
the relative magnification $\delta_{\mu}$:

\begin{equation}
	\delta_{\mu} \equiv \mu - 1 \approx 2\kappa
\label{dmu_eq}
\end{equation}

where the latter approximation holds in the weak lensing limit.

One typically measures magnification by comparing properties of lensed objects to those of 
a large sample.  In this analysis, we measure the change in average magnitude of sources behind lenses 
as well as the change in detection rate (counts) of the sources.

\subsection{Magnitudes}

The apparent magnitude of a single object $i$ is changed by magnification according to:

\begin{equation}
	\delta_{m_{i}}(\mu) = -\frac{2.5}{\mathrm{ln}(10)} \delta_{\mu}.
\label{d_mu_i_equ}
\end{equation}

The corresponding average change in magnitude for a population of galaxies 
has an additional term due to the change in the sample's limiting magnitude $m_{*}$ 
in a magnified region (i.e. more galaxies are brought above the magnitude limit by the 
magnification effect):

\begin{align}
        \delta_{\bar{m}}(\mu) & = \left( \frac{d \bar{m}}{ d m_{i}} \frac{ d m_{i}}{d \mu} + \frac{d \bar{m}}{ d m_{*}} \frac{ d m_{*}}{d \mu} \right) \ \delta_{\mu} \\
        & = -\frac{2.5}{\mathrm{ln}(10)} (1 - \frac{d \bar{m}}{dm_{*}}) \ \delta_{\mu}
\end{align}

where in the last equality we have used equation \ref{d_mu_i_equ}.  The change in magnitude of a lensed sample is then 

\begin{equation}
	\delta_{m}(\mu) = \alpha_{m} \delta_{\mu}
\label{mags0_eq}
\end{equation}

where we have introduced $\alpha_{m}$ as 

\begin{equation}
	\alpha_{m} \equiv \frac{2.5}{\mathrm{ln}(10)} \left( -1 + \frac{d\bar{m}(m_{*})}{dm_{*}} \right).
\label{alpha_m_equ}
\end{equation}

We note that $\alpha_{m}$ is related to $C_{s}$ defined in \cite{menard09} via $\alpha_{m} = (-2.5/\mathrm{ln}(10)) \ C_{s}$, with both 
expressing the change in mean magnitude of a sample due to a variation of the sample magnitude limit.

Combining equations \ref{muapprox_eq}, \ref{kappa_eq}, \ref{dmu_eq}, and \ref{mags0_eq}, we find how 
the surface mass density of the lens can be calculated from $\delta_{m}(\mu)$.  As this surface 
mass density is determined from the lensing-induced fluctuation in magnitudes, we call it $\Sigma_{m}$:

\begin{equation}
	\Sigma_{m} = \frac{\delta_{m}(\mu) \Sigma_{cr}}{2 \alpha_{m}}.
\label{sigmam_eq}
\end{equation}

\subsection{Counts}
\label{counts_section}

The number of detections in a flux-limited survey is altered by gravitational lensing because of two effects.
First, magnification increases the area behind lenses, and this decreases the number of sources:

\begin{equation}
	\delta_{c}(\mathrm{area}) \equiv \frac{dn}{n} = - \delta_{\mu}
\end{equation}

where $n$ is the number density of sources.

Second, magnification also changes the object fluxes, as discussed above.  This causes objects that are intrinsically fainter
than the survey’s magnitude limit $m_{*}$ to appear brighter, and therefore be detected:

\begin{align}
	\delta_{c}(\mathrm{flux}) & \equiv \frac{dn}{d\mu} \frac{\delta_{\mu}}{n} \\ \notag
	& = \frac{2.5}{\mathrm{ln}(10)} \frac{dn}{dm_{*}} \frac{\delta_{\mu}}{n} \\ \notag
	& = -2.5 \frac{d}{dm_{*}}\  (\mathrm{log}_{10}n) \delta_{\mu} \notag
\end{align}

The total lensing-induced change in number counts $\delta_{c}(\mu)$ is the sum of these two components, and so we define

\begin{equation}
	\delta_{c}(\mu) = \delta_{c}(\mathrm{flux}) + \delta_{c}(\mathrm{area}) = \alpha_{c} \delta_{\mu}
\label{counts0_eq}
\end{equation}

where

\begin{equation}
	\alpha_{c} \equiv 2.5 \frac{d}{dm_{*}} \mathrm{log}_{10} n_{\mathrm{o}}(<m_{*}) - 1.0
\label{alpha_c_eq}
\end{equation}

with $n_{\mathrm{o}}$ the unlensed source number density.  $\alpha_{c}$ therefore relates the lensing-induced change in number counts 
$\delta_{c}(\mu)$ to the relative magnification $\delta_{\mu}$.

Analogously to equation \ref{sigmam_eq}, we can define $\Sigma_{c}$, the surface mass density of a 
lens as calculated by the lensing-induced change in counts of a sample of sources:

\begin{equation}
	\Sigma_{c} = \frac{\delta_{c}(\mu) \Sigma_{cr}}{2 \alpha_{c}}.
\label{sigmac_eq}
\end{equation}

\section{Dust Extinction} 
\label{dust_section}

Dust in the lenses absorbs light from the background sources, creating an extinction effect that can be of comparable amplitude with the flux changes 
due to lensing magnification.  Dust mass appears to trace total halo mass \citep{menard09, hildebrandt12}, which causes the two signals to have 
similar radial dependence.  Fortunately, dust extinction is wavelength dependent while gravitational lensing is achromatic, allowing us to use multicolor 
photometry of the sources to disentangle the two effects.  

The change in V-band magnitude of a source due to dust extinction is

\begin{equation}
	\delta_{m_{\tau}(V)} = A_{V} = R_{V} E(B-V).
\label{av_eq}
\end{equation}

The total extinction $A_{V}$ denotes the $V$ magnitude change due to dust extinction.
The relative extinction $E(B-V)$ quantifies how the extinction changes the object's $B-V$ color.  
The two can be related through the parameter $R_{V}$, which depends on the physical properties of the dust such as the typical grain size.  
$R_{V}$ is commonly taken to be about 3, as observed for interstellar dust in 
our Galactic disk and the small and large Magellenic clouds (SMC and LMC, respectively).  
We take $R_{V} = 2.93$, as observed for the SMC.  

The relative extinction $E(B-V)$ is equal to the change in color, $\Delta (m_{B} - m_{V})$.  In fact, we use the color ($m_{g} - m_{i}$) of the sources 
to measure the reddening, as those bands are measured by SDSS and this color has larger signal to noise than the difference between closer filters 
such as $g$ and $r$ or $r$ and $i$.  To relate the observed source color fluctuations to the amount of dust in the lenses, we must convert the observed 
$\Delta (m_{g} - m_{i})$ to $\Delta (m_{B} - m_{V})$ in the lens rest frame.  We do this by assuming the SMC dust model from \cite{pei92}, which 
parameterizes the color dependence of extinction, and by using a template spectrum for the source LRGs.  
As discussed in \cite{pol} and described in section \ref{megaz_section}, BPZ was run on the MegaZ source sample using CWW templates of the LePhare 
library.\footnote[1]{The templates can be found in the folder /lephare.dev/sed/GAL/CE\_NEW/ of the LePhare package at http://www.cfht.hawaii.edu/$~$arnouts/LEPHARE/DOWNLOAD/ lephare\_dev\_v2.2.tar.gz}.  
The majority of the galaxies were matched to the LRG template \texttt{Ell\_01.sed}, which we therefore assume to be representative of the sample.  
With these models we can relate observed ($m_{g} - m_{i}$) to lens rest frame $(m_{B} - m_{V}$), to 
absolute extinction $A_{V}$ in the lens frame, and to observed extinction $A_{i}$.

To relate the extinction in the lens, $A_{V}$, to the total lens surface mass density $\Sigma$, we must assume the opacity of the dust 
$K_{\mathrm{ext}}$ that relates the dust's absorption cross section to its mass and depends on the dust grain composition.  
We take $K_{\mathrm{ext}} = 1.54 \times 10^{4} \mathrm{cm}^{2} \mathrm{g}^{-1}$, 
following \cite{menard09}, from the SMC dust model in \cite{weingartner01}.
If we assume a constant dust-to-mass ratio $\Gamma$ in the lenses, then 

\begin{equation}
	\Sigma = \frac{ \mathrm{ln}(10) }{2.5} \frac{A_{V}}{\Gamma \times K_{\mathrm{ext}} }.
\label{dtm_eq}
\end{equation}

\section{Observables}

Both lensing and dust extinction contributions must be taken into account when interpreting 
the measured fluctuations in object counts and magnitudes.

\subsection{Counts}

The fluctuation in observed counts due to lensing and extinction in a flux-limited survey can be written, 
in an elaboration of equation \ref{counts0_eq}, as 

\begin{equation}
	\delta_{c} \equiv \frac{n}{\bar{n}} - 1 = \alpha_{c} \delta_{\mu} - \frac{ \mathrm{ln}(10) }{2.5} (\alpha_{c}+1) A_{i}.
\label{counts_eq}
\end{equation}

The factor $(\alpha_{c}+1)$ relates the magnitude change in $m_{*}$ to the observed number of counts, 
while the term $\mathrm{ln}(10)/2.5$ results from the derivative of the base 10 logarithm 
in $\alpha_{c}$.  The subscript $i$ in the extinction $A_{i}$ refers to the fact that we measure changes in 
sources' i-band magnitudes.

Here $n$ and $\bar{n}$ are the observables: $n$ is the number count density of objects in a given region, 
while $\bar{n}$ is the mean number density of counts over the sky, 
which is used as an observable approximation to the unlensed number count density $n_{\mathrm{o}}$.

\subsection{Magnitudes}

The average magnitude of a set of galaxies, compared to the magnitude averaged over the whole sky, can be written as:

\begin{equation}
	\delta_{m} \equiv m - \bar{m} = \alpha_{m} \delta_{\mu} -  \frac{ \mathrm{ln}(10) }{2.5} \alpha_{m} A_{i}.
\label{mags_eq}
\end{equation}

Because extinction and lensing both affect only the measured flux (i.e. there is no geometrical contribution 
as in the case of number counts), $\alpha_{m}$ affects both terms on the right hand side in the same way.  The factor of 
$\frac{ \mathrm{ln}(10) }{2.5}$ appears as a conversion factor because $A_{i}$ is in units of magnitudes while $\delta_{\mu}$ 
is unitless.

\subsection{Conversion to Surface Mass Density}

The surface mass density of the lens, taking into account both lensing and extinction, is analogous 
to equations \ref{sigmam_eq} and \ref{sigmac_eq} and has the same form in the case of counts and magnitudes 
observables:

\begin{equation}
	\Sigma = \frac{ \delta_{c,m} \Sigma_{cr} }{2 \alpha_{c,m} }.
\label{sigma_eq}
\end{equation}

This implies that the surface mass density $\Sigma$ can be obtained by applying a dust correction to $\Sigma_{m}$ or 
$\Sigma_{c}$:

\begin{equation}
	\Sigma = \Sigma_{m} - \frac{ \mathrm{ln}(10) }{2.5} \frac{ A_{i} \Sigma_{cr} }{2}
\label{sigmam2sigma_eq}
\end{equation}
\begin{equation}
	\Sigma = \Sigma_{c} - \frac{ \mathrm{ln}(10) }{2.5} \frac{\alpha_{c}+1}{\alpha_{c}} \frac{ A_{i} \Sigma_{cr} }{2}
\label{sigmac2sigma_eq}
\end{equation}

\section{Lens Modeling}

\subsection{Main Halo}

We model the dark matter halo, comprising the majority of the mass in the lens, as an NFW profile \citep{nfw}:

\begin{equation}
\rho(r) = \frac{\delta_{c} \rho_{\mathrm{crit}}}{(r/r_{s})[1+(r/r_{s})]^{2}}
\label{nfw_equ}
\end{equation}
where $c$ is the concentration parameter of the profile, $\delta_{c} = \frac{200}{3} \frac{c^{3}}{\mathrm{ln}(1+c) - c/(1+c)}$, and $r_{s} = R_{200}/c$.  The total mass inside $R_{200}$ is $M_{200}$.  $\rho_{\mathrm{crit}}$ is the critical density of the universe at the redshift of the cluster.

The concentration of haloes has been seen to vary inversely with mass and redshift.  We assume the relation measured in \cite{mandelbaum08}:
\begin{equation}
c(M_{200},z) = \frac{4.6}{1+z} \left(\frac{M_{200}}{1.56 \times 10^{14} h^{-1} M_{\odot}}\right)^{-0.13}.
\label{m_c_eq}
\end{equation} 

The surface mass density profile $\Sigma$ of an NFW lens can be calculated analytically, as given in \cite{schneider04}.

We take $M_{200}$, the mass of the halo, to be the single free parameter in the model for the main halo.

\subsubsection{Halo Miscentering}

The centers of massive haloes are difficult to identify.  Typically, the stated center of a galaxy cluster is chosen as the location of the brightest 
cluster galaxy (BCG).  
In some cases, the BCG may be offset from the true halo center.  More often, the BCG is misidentified.  Such errors in the determination of the halo 
center will lead to errors in the stacked halo profile, in particular making the stacked profile shallower than expected.  
\cite{johnston07} studied this problem in detail using simulations, resulting in a mass-dependent probability that a cluster is offset from its 
true center by a given amount.  We implement this prescription in our NFW model of cluster haloes by using a 
composite halo profile that includes a contribution from miscentered haloes according to the probability and radial offsets given in \cite{johnston07}.  
This contribution makes the halo profile shallower, although not significantly given the measurement errors in our data.  

We note that the miscentering prescription we implement is calibrated to match the behavior of a cluster catalog with a different selection 
technique from \cite{dr9sz}.  Furthermore, there are hints that the modeled miscentering is overestimated \citep{mandelbaum08}.  
We use the prescription as an example of the effects of miscentered haloes in the catalog, since we do expect a significant level of miscentering 
in any cluster sample, particularly when the center is identified as the location of the BCG as in both \cite{johnston07} and \cite{dr9sz}.  
We examine the effects of including miscentering in our model in section \ref{miscentering_section}.

\subsection{BCG}

We model the brightest cluster galaxy (BCG) of the halo as a cuspy profile defined similarly to equation \ref{nfw_equ}:
\begin{equation}
\rho(r) = \frac{\delta_{c} \rho_{\mathrm{crit}}}
{(r/r_{s})^{\gamma}[1+(r/r_{s})^{2}]^{(n-\gamma)/2}}
\label{cuspy_equ}
\end{equation}
with $(\gamma, n) = (2,4)$.  This profile has steeper inner and outer slopes than the NFW profile, which is similar to equation \ref{cuspy_equ} with 
$(\gamma, n) = (1,3)$.  We note that in lensing shear analyses of massive haloes the BCG has typically been modeled simply as a point mass 
\citep[e.g.][]{johnston07}.  This is unfeasible in the case of magnification modeling, where the observable is related to $\kappa$ rather than the change in 
$\kappa$.

In our model, we fit for $M_{200}$ of the BCG halo profile;  as with the NFW profile, $M_{200}$ is the mass enclosed in $R_{200} = r_{s} c$.  
For the BCG we again use the mass-concentration relation given in equation \ref{m_c_eq}.

\subsection{Second Halo Term}

At large radii we expect to see lensing signal from haloes neighboring the main one.  To calculate this second halo term we follow the procedure in 
\cite{mandelbaum05} to calculate the power spectrum of the second halo term using the NFW profile (whose Fourier transform $y_{\mathrm{dm}}$ is given 
analytically in \citealt{scoccimarro01}), the linear power spectrum $P_{\mathrm{lin}}$ (as calculated by the code 
NICAEA\footnote[2]{http://www2.iap.fr/users/kilbinge/nicaea/}), and a bias $b$ of the lens:

\begin{equation}
	P(k) = b P_{\mathrm{lin}}(k,z) \int f(\nu) \mathrm{d}\nu b(\nu)  y_{\mathrm{dm}}(k, M)
\end{equation}

The functions $b(\nu)$ and $f(\nu)$ are the bias and mass functions derived using the spherical collapse formalism 
(following \citealt{mandelbaum05}; see \citealt{press_schechter,sheth_tormen}).

The Fourier transform of $P(k)$ yields $\xi_{\mathrm{nfw,dm}}$, the cross-correlation between the lens halo and the neighboring dark matter.  Integrating 
along the line of sight produces the projected mass density $\Sigma$ of the second halo term:

\begin{equation}
	\Sigma(R) = \bar{\rho} \int \xi_{\mathrm{nfw,dm}} [(R^{2} + \chi^{2})^{1/2}] d \chi
\end{equation}

where $\bar{\rho}$ is the mean density of the universe at the redshift of the halo.

We fit for the lens bias $b$ as the free parameter in the second halo term.

\section{Data}
\label{data_section}

\subsection{Sources}
\label{megaz_section}

\begin{figure}
\includegraphics[width=85mm]{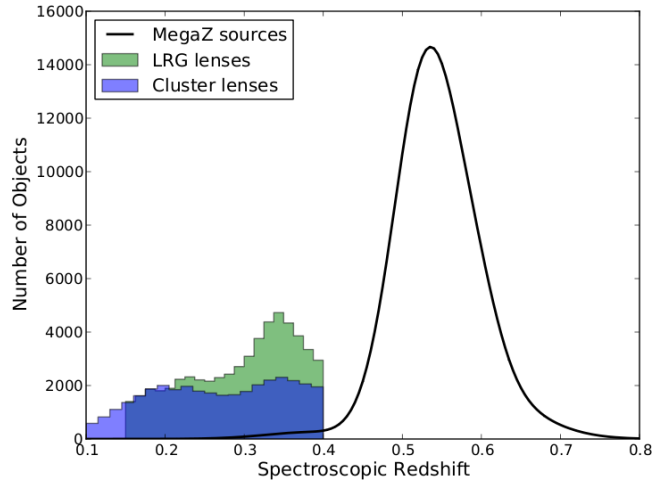}
\caption{Redshift distributions of the LRG lenses (green), cluster lenses (blue), and MegaZ sources (black).  The MegaZ curve is a fit using a subset of the data with spectroscopic redshifts;  see text for details.}
\label{nofz_fig}
\end{figure}

Our source sample is the MegaZ catalog of luminous red galaxies (LRGs) \citep{megazdr7} color-selected from the Sloan Digital Sky Survey 
(SDSS) Data Release 7 \citep{sdss7}.  The catalog contains 1,120,745 LRGs with photometric redshifts between 0.4 and 0.8, over a sky area 
of 7700 square degrees.  For this 
study we use a subset of 572,900 galaxies with photometric redshift between 0.5 and 0.6 with deVaucouleurs $i$ magnitude brighter than 19.8.  
We can quantify the accuracy of the photometric redshifts of this subsample because the galaxies have the same selection criteria as the 
spectroscopic 2SLAQ catalog \citep{2slaq}.  Precise understanding of the redshift distribution N(z) of the source sample is essential, 
as physical overlap between the sources and the lenses leads to a spurious lensing signal (see section \ref{overlap_section}).  
We use photometric redshifts for the sample calculated using the BPZ code \citep{bpz} as discussed in \cite{pol}.  
The N(z) of the photo-z bin 0.5-0.6 can be fit well by a sum of four Gaussian distributions, in order to parameterize accurately the shape of 
the peak and both high and low redshift tails.  
Fitting 3 parameters per Gaussian distribution, including covariance between the 12 parameters, yields a $\chi^{2}$ = 23.4 for 18 degrees of freedom; 
the best fit is shown in black in Figure \ref{nofz_fig}.

We use the deVaucouleurs $i$ magnitude to measure $\delta_{m}$, and the model magnitudes $g$ and $i$ to measure color.  The SDSS data, as 
published in the MegaZ catalog, have been calibrated to the 2\% level\footnote[4]{http://www.sdss.org/dr7/algorithms/fluxcal.html};  we therefore 
add a systematic uncertainty of 2\% to the magnitude and color measurements, on top of the jackknife uncertainties that constitute our 
primary error analysis.  This is typically insignificant for the $\Sigma_{m}$ measurements, but is large compared to the amplitude of the measured reddening 
of the sources.

\begin{figure}
\begin{center}
\includegraphics[width=85mm]{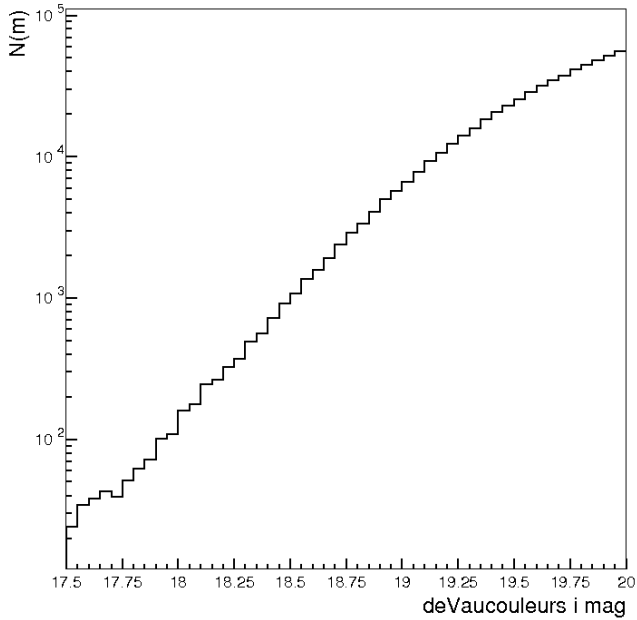}
\end{center}
\caption{Number of MegaZ galaxies with redshift 0.5-0.6, versus deVaucouleurs $i$ mag.  For the analysis the cut is chosen at $i_{deV}$ = 19.8.}
\label{n_m_fig}
\end{figure}

\begin{figure}
\begin{center}
\includegraphics[width=85mm]{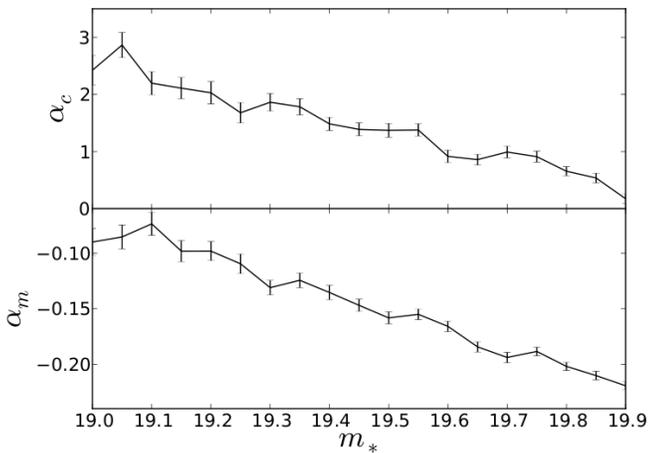}
\end{center}
\caption{$\alpha_{c}$ (top) and $\alpha_{m}$ (bottom) versus the limiting magnitude cut $m_{*}$ on the MegaZ catalog.  For the analysis the cut is chosen at $i_{deV}$ = 19.8.}
\label{alpha_fig}
\end{figure}

The number of source galaxies versus deVaucouleurs $i$ magnitude is shown in Figure \ref{n_m_fig}.  
Measurements of $\alpha_{c}$ and $\alpha_{m}$ for the MegaZ sample are shown in Figure \ref{alpha_fig}, versus the chosen limiting magnitude cut 
in de Vaucouleurs $i$.  For the analysis we choose $i<19.8$ because this corresponds to the faint limit of the 2SLAQ photometric redshift 
calibration sample;  the MegaZ catalog extends to $i=20$.  The distribution of the catalog number counts versus 
magnitude is modeled using kernel density estimation.  The uncertainties on $\alpha_{c,m}$ are fundamentally uncertainties on the histogram shown in Figure 
\ref{n_m_fig}, which can be caused by Poisson error, magnitude measurement error, or intrinsic variations in the number of objects per magnitude across 
the sky due to the presence of large scale structure.  We estimate these errors on $\alpha_{c,m}$ by jackknife resampling in 133 regions over the 
survey area.  
In order to incorporate these measurement errors into our analyses, we include $\alpha_{c}$ and 
$\alpha_{m}$ as fit parameters in our analysis but constrain them with Gaussian priors around the best-fit values of 0.656 and -0.219, 
with 1$\sigma$ errors of 12\% and 2\%, respectively.  Note that $\alpha_{m}$ is much better constrained than $\alpha_{c}$ due to the 
smaller intrinsic correlation of galaxy magnitudes compared to galaxy positions.

\subsection{Lenses}

\subsubsection{Cluster Catalog}

We use the catalog of galaxy clusters presented in \cite{dr9sz}, which are selected using photometric redshift measurements from SDSS data release 
3 \citep{sdss3}.  For our cluster lens sample we take the subset of 28,617 clusters with spectroscopic redshift 
identifications\footnote[5]{http://zmtt.bao.ac.cn/galaxy\_clusters/} $z_{spec}$ between 0.1 and 0.4.  The redshift distribution is shown in 
blue in Figure \ref{nofz_fig}.  We divide the sample into 5 bins of richness, shown as bins 1-5 in Table \ref{richness_table}.

\begin{table}
\begin{center}
\begin{tabular}{|l|l|l|}
\hline
Richness Bin &  Richness $N$ & \# Lenses \\
\hline
0 & $<$12 & 66613 \\
1 & 12-17 & 12270 \\
2 & 18-25 & 9765 \\
3 & 26-40 & 4649 \\
4 & 41-70 & 1638 \\
5 & 71-220 & 295 \\
\hline
\end{tabular}
\end{center}
\caption{Richness bin limits, and number of lenses per bin.  Bins 1-5 constitute the cluster sample, while bin 0 corresponds to the spectroscopic LRG sample.}
\label{richness_table}
\end{table}

\subsubsection{LRGs}

We also study a spectroscopic sample of luminous red galaxies (LRGs) identified in SDSS.  This is the sample selected in \cite{cabre09} using 
magnitude, color, and surface brightness cuts from SDSS data release 6 \citep{sdss6}.  
Of the 73,981 LRGs with redshifts between 0.1 and 0.4, 10\% match with cluster catalog members within a radius of 10".  
We use as our LRG lens 
sample the remaining 66,613 galaxies and list them as richness bin 0 in Table \ref{richness_table}, with richness 
less than 12.  
Their spectroscopic redshift distribution is shown in green in Figure \ref{nofz_fig}.

\subsection{Spatial Mask}
\label{mask_section}

When calculating spatial fluctuations in the source detection rate, it is essential to use an accurate 
mask of the survey footprint.  We generate a mask in Healpix format \citep{healpix} with resolution 
Nside=8192, corresponding to pixel size of 26 arc seconds.  When counting detections in an annulus around a lens, we query 
the mask for pixel centers within the annulus and correct our counts by the fraction of masked pixels in the 
annulus area.  The relevant mask for the analysis is that of the MegaZ sources, as those are the objects for which 
we calculate the average density and spatial fluctuations.  The LRG and cluster catalog cover similar footprints 
to the MegaZ, as all three are drawn from SDSS data.  The cluster catalog includes additional area in the 
southern galactic cap that is not used in this analysis, as it does not overlap with the MegaZ footprint.

\begin{figure}
\begin{center}
\includegraphics[width=85mm]{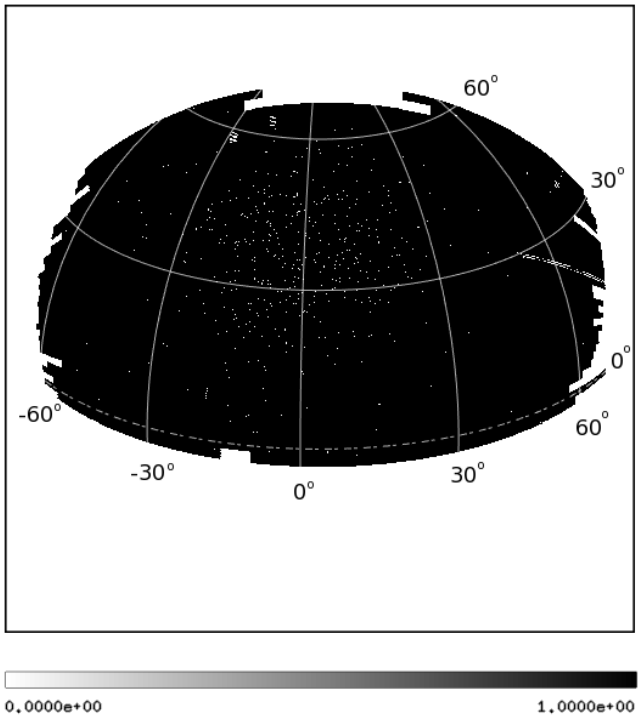}
\end{center}
\caption{Mask used in the analysis, as described in Section \ref{mask_section}.}
\label{mask_fig}
\end{figure}

We calculate the mask for the MegaZ catalog in a two-step process.  First, we use the Healpix mask for the 
sample, with $N_{side}$=1024, that is provided with the catalog.  This mask provides the overall 
footprint of the survey.  Second, we mask regions of poor SDSS data quality inside the footprint by 
examining the density of stars in the area.  We download from the SDSS CasJobs 
server\footnote[6]{http://casjobs.sdss.org/CasJobs/} all stars in the region 
brighter than $r$ magnitude 19.6.  A map of this dense stellar catalog clearly shows regions with systematically 
lower density, likely due to poor observing conditions.  We generate a star map with $N_{side}$=512 and mark pixels 
as bad if they include fewer than 8 stars.  This cut removes areas such as a long line in the SDSS driftscan 
direction at high right ascension and intermediate declination, 
as well as some rectangular regions at high declination and negative 
right ascension.  The cut also eliminates small disjoint regions at high galactic latitudes that 
have few stars due to statistical fluctuations.  As these regions are spatially uncorrelated with the positions 
of gravitational lenses, masking them will not bias our analysis.  The intersection of the resulting 
star mask with the MegaZ footprint mask is upgraded to $N_{side}$=8192 for use in our analysis, 
and is shown in Figure \ref{mask_fig}.

\section{Method}

To maximize our signal to noise in measuring the average mass profiles of the lenses, we stack the lens galaxies and clusters in 
bins of richness.  The spectroscopic LRG sample is stacked all together, while the cluster catalog is divided into five richness bins, 
as given in Table \ref{richness_table}.

We look around each lens in logarithmic radial bins.  
In these radial annuli we measure the source counts, average magnitude, and average color, 
and compare them to the means from the entire source catalog.  Using the measured fluctuations in magnitude ($\delta_{m}$) and 
counts ($\delta_{c}$), combined with the $\alpha$s measured 
as described above, we calculate $\Sigma_{m}$ and $\Sigma_{c}$ via equations \ref{sigmam_eq} and \ref{sigmac_eq}.  

The change in color $g-i$ in annular bins around the lens, compared to the average $g-i$ of the MegaZ LRGs, constitutes our 
measurement of the reddening E(g-i).  This reddening profile can be converted to a measurement of the halo mass profile 
using equations \ref{av_eq} and \ref{dtm_eq}, with the dust-to-mass fraction $\Gamma$ as a free parameter.  

Our model consists of an NFW profile, BCG profile, a second halo term, and a dust-to-mass ratio.  The sum of the first three components 
yields the total mass density $\Sigma$ of the lens.  The dust-to-mass ratio $\Gamma$ then relates this $\Sigma$ to $A_{i}$ and E(g-i) 
as described in Section \ref{dust_section}.  This model E(g-i) is compared directly to our reddening measurements.  The model 
$A_{i}$ can be used to calculate models for $\Sigma_{c}$ and $\Sigma_{m}$ using equations \ref{sigmam2sigma_eq} and \ref{sigmac2sigma_eq}, 
which are compared directly with our measurements of the counts and magnitudes fluctuations.

Measurement errors are determined using jackknife resampling over 133 regions across the survey area.  We measure the 
covariance across radial bins 
and also between the counts, magnitudes, and reddening measurements.  The different richness subsamples are fit independently.  

Because our source and lens catalogs span a range of redshifts, $\Sigma_{cr}$ varies by a factor of nearly 2 over the data set.  If 
we approximate the $\Sigma_{cr}$ of the sample by its mean, we will smear our results when translating from $\delta_{\mu}$ to 
$\Sigma$, the projected mass density of the lens.  We therefore divide the analysis into redshift sub-bins of width roughly 0.05, which 
reduces the difference in $\Sigma_{cr}$ between neighboring bin pairs to typically 10\%.  $\Sigma$ is calculated separately for each 
pair of foreground/background bins.  The final result is obtained by averaging them, weighted by the number of measurements divided by 
$\Sigma_{cr}^{2}$.

Our model for the lenses includes six free parameters: $M_{200}$ of the NFW lens halo profile, $M_{200}$ of the cuspy BCG profile, bias $b$ 
as determined through the second halo term, dust-to-mass fraction $\Gamma$, $\alpha_{c}$ and $\alpha_{m}$.  The parameters have flat priors except for 
the $\alpha$s, which have Gaussian priors with central values and 1$\sigma$ errors given by the measurements described in section 
\ref{megaz_section}.  We explore the likelihood using Markov Chain Monte Carlo (MCMC) sampling, fitting each richness bin independently.

\section{Results}

\begin{figure}
\includegraphics[width=85mm]{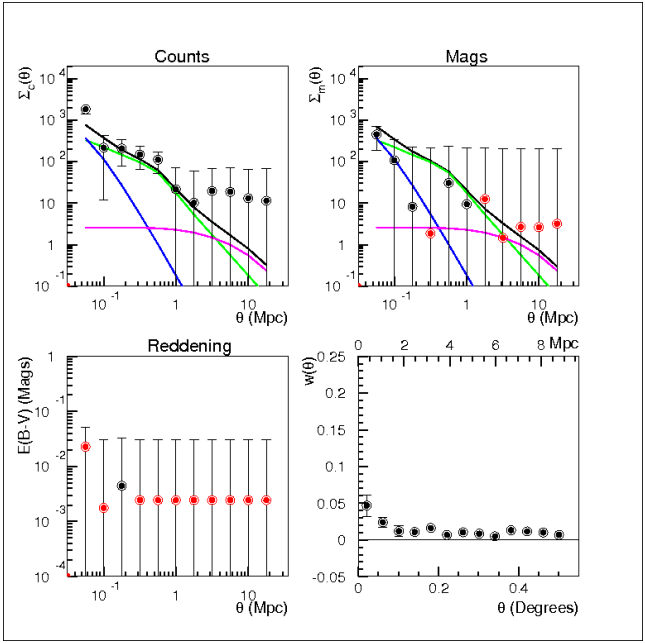}
\caption{Magnification measured using counts (top left) and magnitudes (top right) for the LRG sample, with units $\mathrm{M_{\odot}}/\mathrm{pc}^{2}$.  Reddening is shown in the bottom left, in units of magnitudes.  The cross-correlation between the lenses and sources, in angular bins with units of degrees is shown in the bottom right, with the top-x axis showing the physical scale at the central redshift of the lenses for reference. }
\label{halo_lrgs_fig}
\end{figure}

\begin{figure}
\begin{center}
\includegraphics[width=85mm]{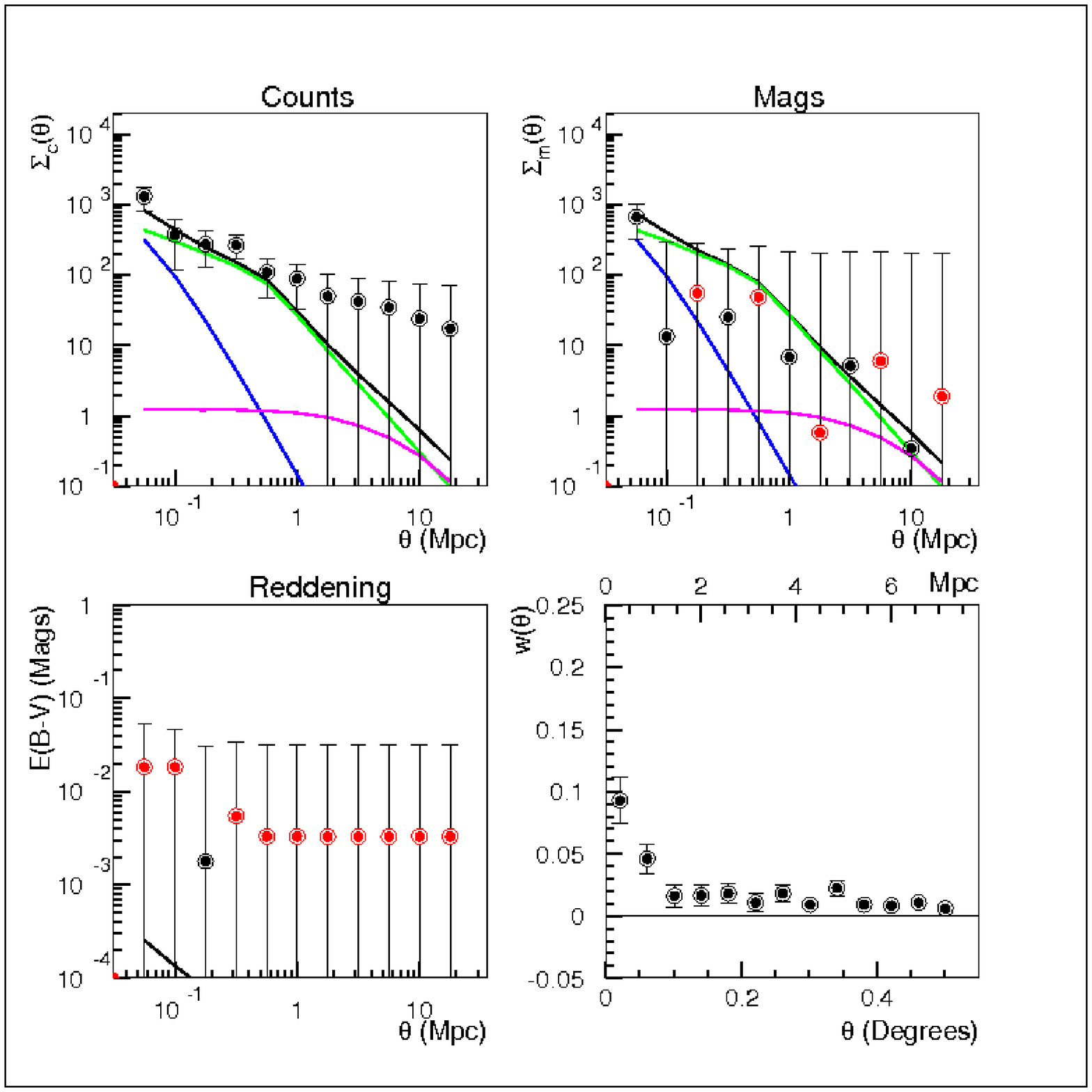}
\end{center}
\caption{As in Figure \ref{halo_lrgs_fig}, for the cluster lens sample with richness 12-17.}
\label{halo_n1_fig}
\end{figure}

\begin{figure}
\begin{center}
\includegraphics[width=85mm]{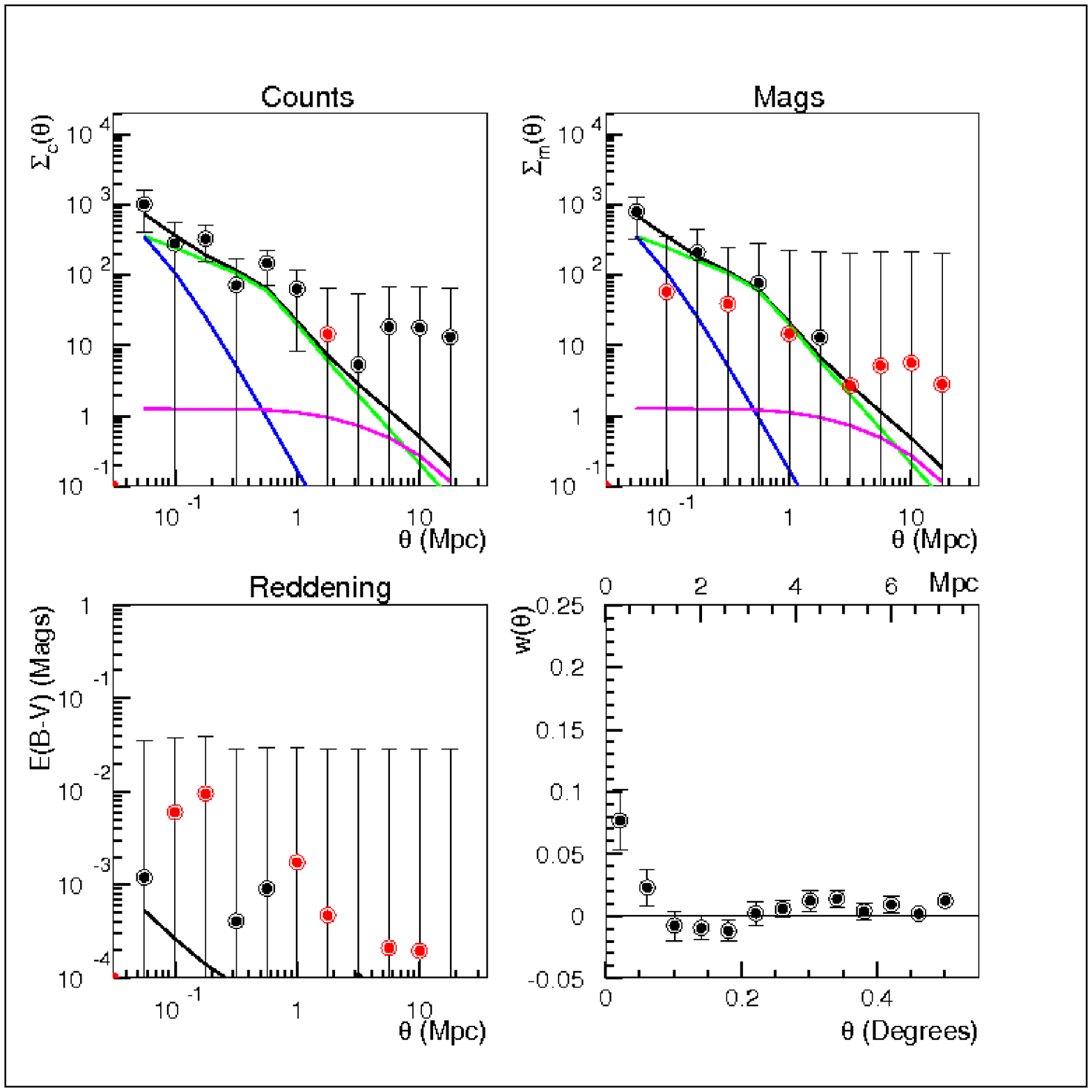}
\end{center}
\caption{As in Figure \ref{halo_lrgs_fig}, for the cluster lens sample with richness 18-25.}
\label{halo_n2_fig}
\end{figure}

\begin{figure}
\begin{center}
\includegraphics[width=85mm]{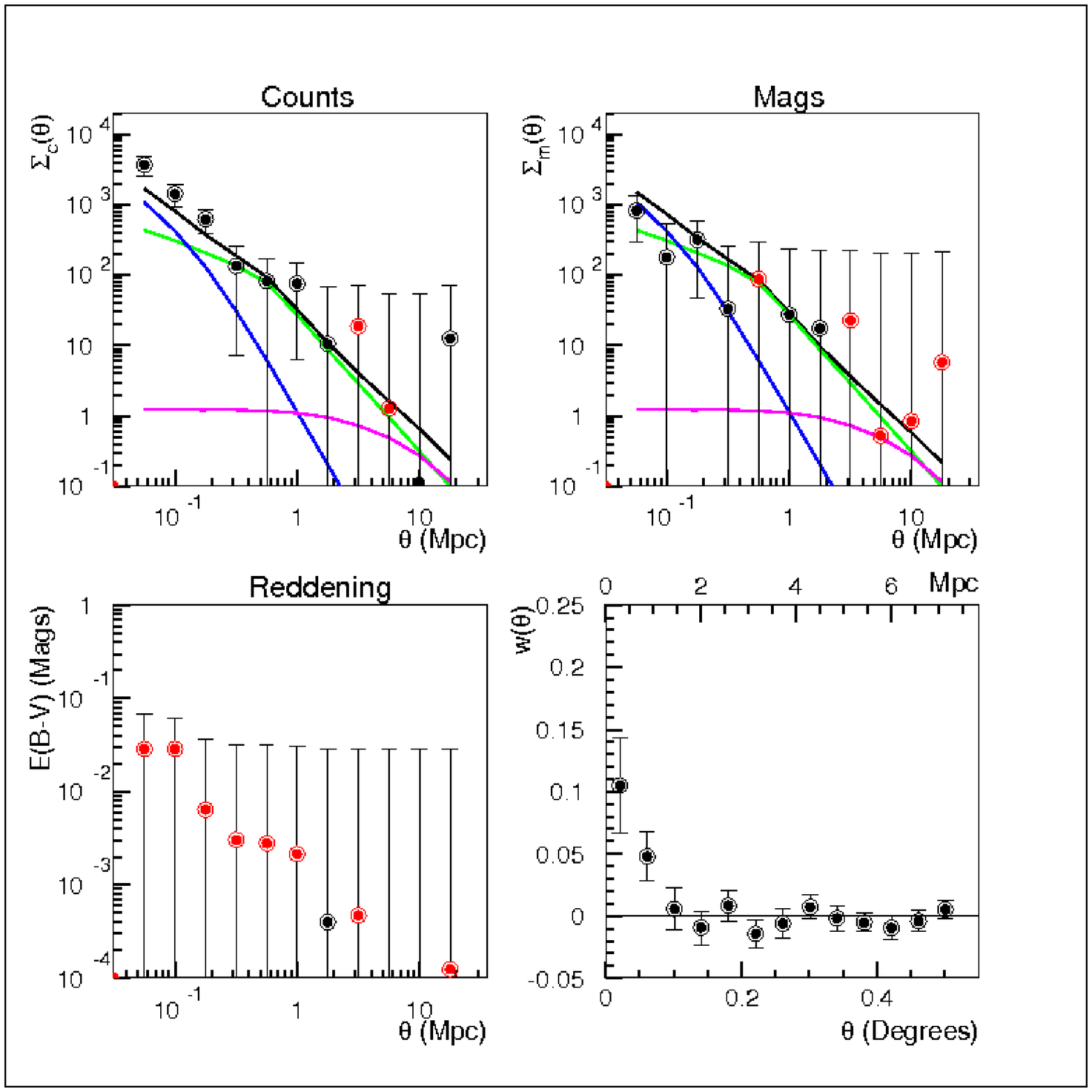}
\end{center}
\caption{As in Figure \ref{halo_lrgs_fig}, for the cluster lens sample with richness 26-40.}
\label{halo_n3_fig}
\end{figure}

\begin{figure}
\begin{center}
\includegraphics[width=85mm]{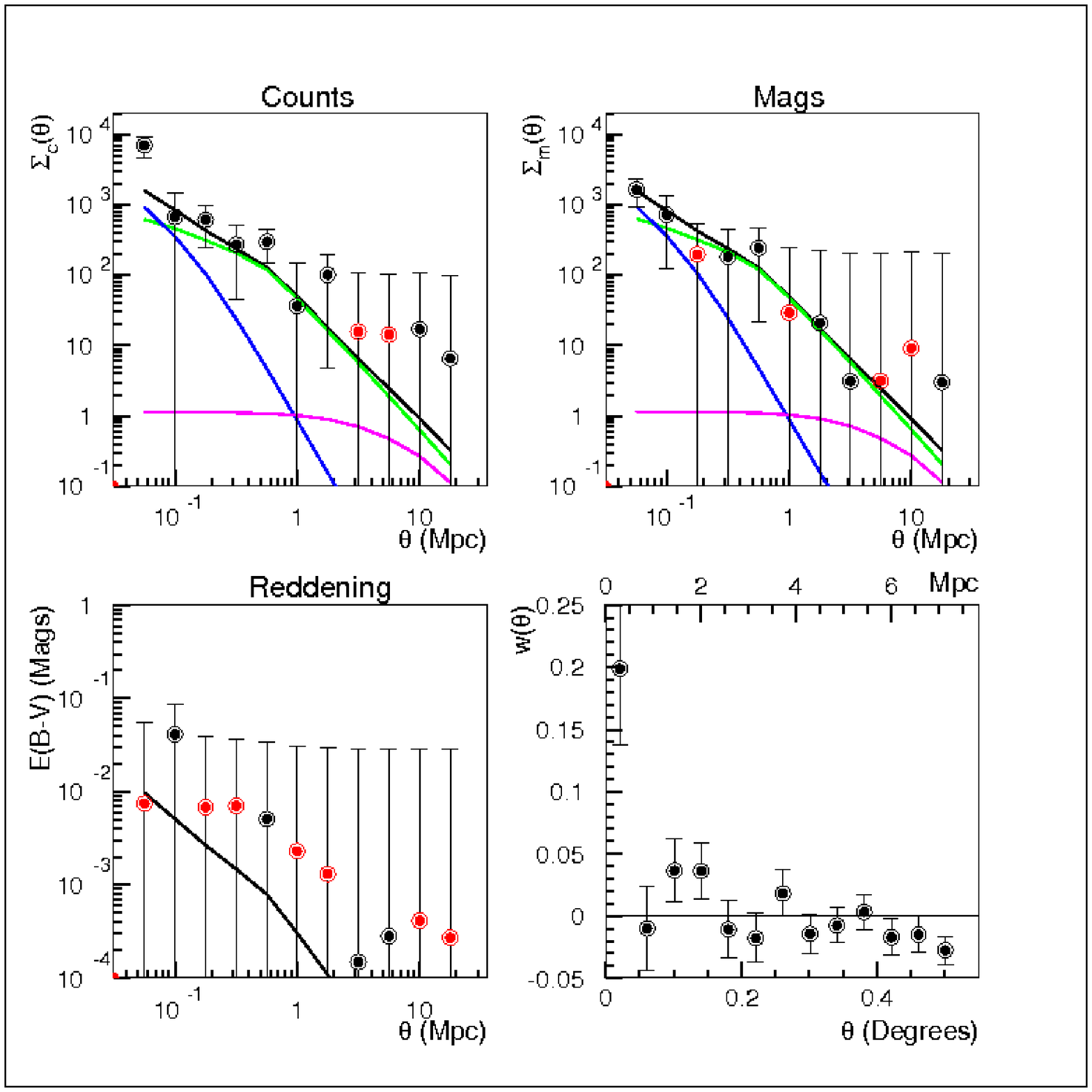}
\end{center}
\caption{As in Figure \ref{halo_lrgs_fig}, for the cluster lens sample with richness 41-70.}
\label{halo_n4_fig}
\end{figure}

\begin{figure}
\begin{center}
\includegraphics[width=85mm]{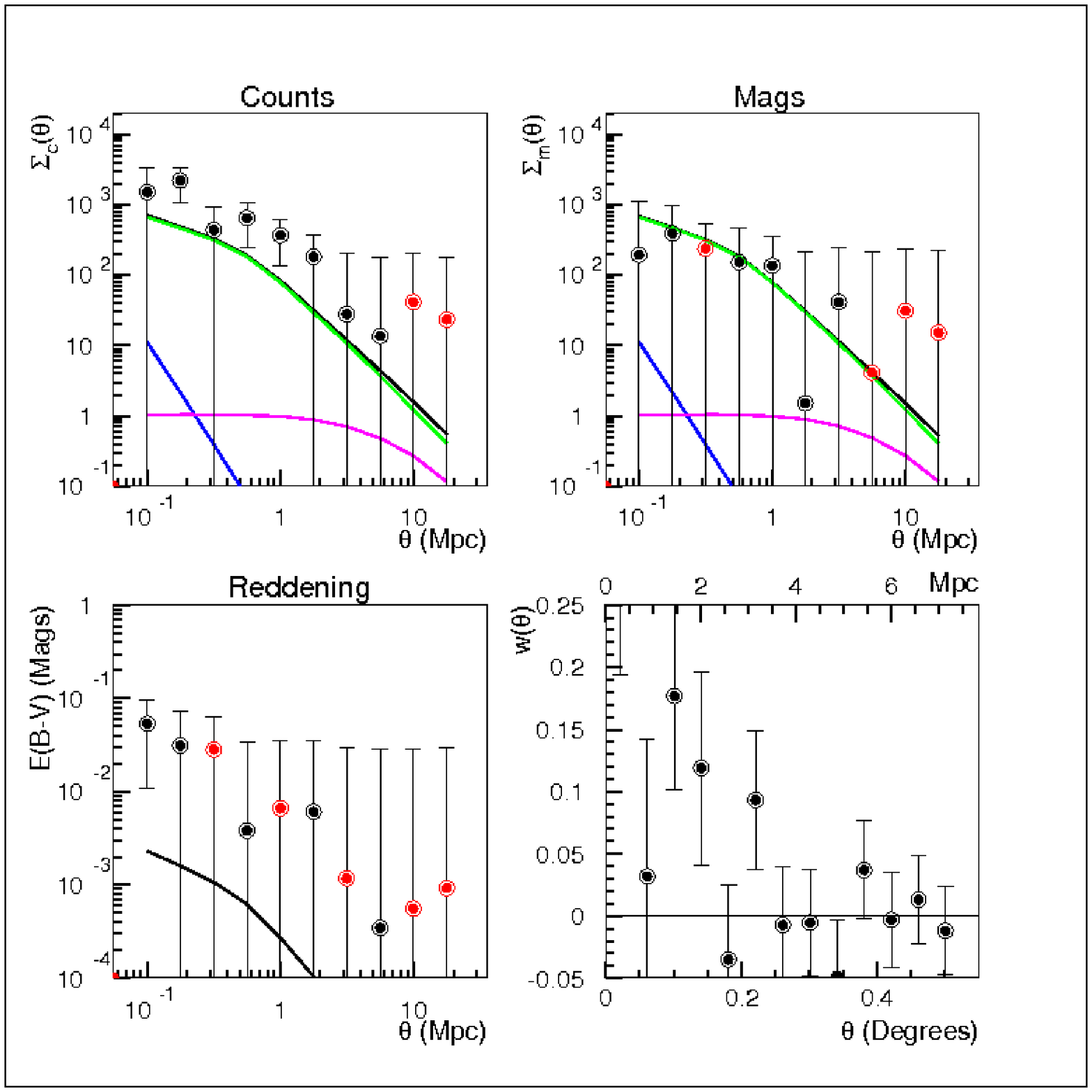}
\end{center}
\caption{As in Figure \ref{halo_lrgs_fig}, for the cluster lens sample with richness 71-220.}
\label{halo_n5_fig}
\end{figure}

The measured projected mass density profiles $\Sigma_{c}$ and $\Sigma_{m}$ are shown in 
Figures \ref{halo_lrgs_fig} through \ref{halo_n5_fig} for the LRG 
sample and the five cluster richness bins, with the 
counts, magnitudes, and reddening measurements shown separately.  The counts and magnitudes results (top left and right, respectively) show the projected mass density profiles 
$\Sigma_{c,m}$ in units of $\mathrm{M_{\odot}}/\mathrm{pc}^{2}$ (without the correction for dust extinction), while the reddening measurements (bottom left) 
show the average E(B-V) in the lens rest frame, calculated from the source model magnitudes $g-i$, the SMC dust model from \cite{pei92}, and the LRG 
template \texttt{Ell\_01.sed}.  
The black data points signify positive values, the red signify 
negative.  
The best-fit profiles for the NFW and BCG components are shown in green and blue; the second halo term is shown in purple;  the sum of these 
components is in black.  Because our data do not 
constrain the second halo term well we fix the lens bias, and therefore the second halo term amplitude, as described in Section \ref{twohalosection}.  
The effects of dust extinction on the counts and magnitudes have been included in the model curves.  
These corrections decrease the model $\Sigma_{c,m}$, with a stronger effect on $\Sigma_{m}$ than on $\Sigma_{c}$.  Note that this implies that not only a 
measurement of reddening, but also a difference in the measurements of $\Sigma_{c}$ and $\Sigma_{m}$ can constrain $\Gamma$. 

The bottom-right plots show the counts cross-correlation between the lens and source catalogs, in angular bins, 
in order to qualitatively compare with the stacking results.  Because the catalogs 
cover a range of redshifts, the correlation function mixes physical scales in its angular bins, causing smaller signal to noise in the 
cross-correlations than in the stacked results.
The x-axis shown on the top of the plot demonstrates, for reference, the scale in Mpc at the central redshift of the lens bin.  We measure a 
significant lensing signal in the inner angular bins that increases with lens richness.

The best-fit model values and 68\% errors are given in Table \ref{results_table} for the NFW and BCG masses 
and the dust-to-mass ratio $\Gamma$.

\begin{table}
\begin{center}
\begin{tabular}{| >{$}l<{$} | >{$}l<{$} | >{$}l<{$} | >{$}l<{$} |}
\hline
\mathrm{Bin} & \mathrm{M_{NFW}\ (\mbox{\sc{e}}14\ M_{\odot})} & \mathrm{M_{BCG}\ (\mbox{\sc{e}}13\ M_{\odot})} &  \mathrm{\Gamma\ (\mbox{\sc{e}-}6)} \\
\hline
0 & 1.26^{+030}_{-0.65} & 1.48^{+0.91}_{-0.79} & 3.08^{+5.73}_{-3.08} \\
1 & 1.95^{+0.55}_{-0.82} & 1.34^{+1.02}_{-0.76} & 3.76^{+6.85}_{-3.76} \\
2 & 1.35^{+0.54}_{-0.76} & 1.60^{+1.33}_{-0.97} & 6.66^{+6.38}_{-6.66} \\
3 & 2.57^{+0.83}_{-1.55} & 3.79^{+2.63}_{-1.87} & 3.30^{+2.87}_{-3.30} \\
4 & 4.07^{+2.02}_{-2.32} & 3.02^{+4.16}_{-1.92} & 5.94^{+6.66}_{-5.94} \\
5 & 5.13^{+2.97}_{-4.13} & 7.82^{+5.03}_{-4.52} & 8.14^{+8.85}_{-6.02} \\
\hline
\end{tabular}
\end{center}
\caption{Best-fit model parameters with 68\% errors.  Richness bin, NFW $M_{200}$, BCG $M_{200}$, dust-to-mass ratio $\Gamma$.  }
\label{results_table}
\end{table}

\begin{figure}
\begin{center}
\includegraphics[width=85mm]{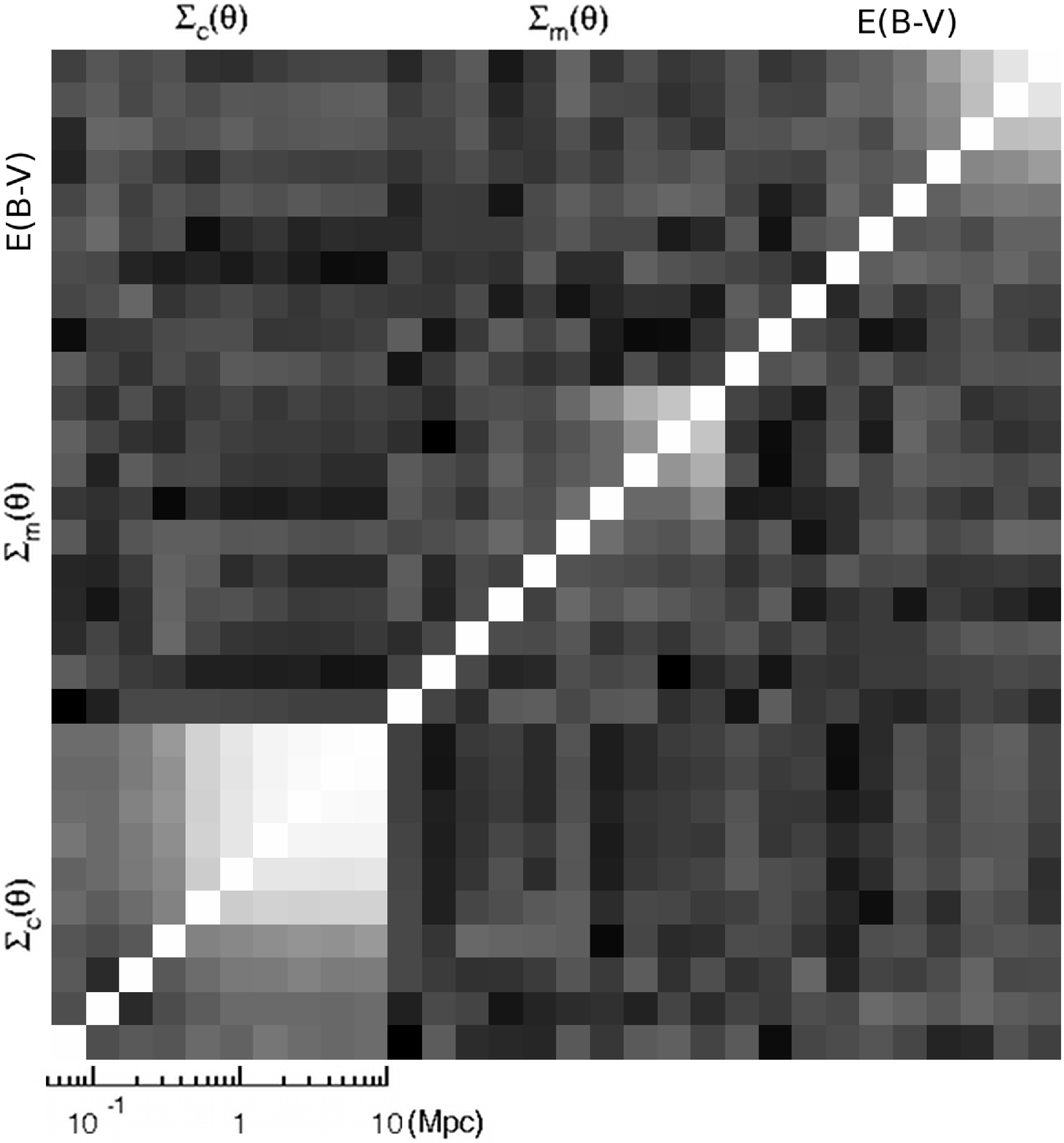}
\end{center}
\caption{Correlation matrix for cluster richness bin 2.  The three blocks are for counts, magnitudes, and reddening (from left to right and bottom to top).}
\label{cov_fig}
\end{figure}

The correlation matrix for lens richness bin 2 is shown as an example in Figure \ref{cov_fig}.  The matrix contains three main blocks, from left to right 
(and bottom to top) corresponding 
to the measurements for counts, magnitudes, and reddening.  
The counts, magnitudes, and color measurements are not significantly cross-correlated, indicating that the physical 
correlation between galaxy luminosity, color, and environment contributes negligible covariance in our analysis.

Note that the correlation between angular bins is much smaller for the magnitudes measurements than for the counts, because the intrinsic 
spatial correlation of magnitudes has smaller amplitude than that of counts.

\subsection{NFW Halo masses}

\begin{figure*}
\includegraphics[width=\textwidth]{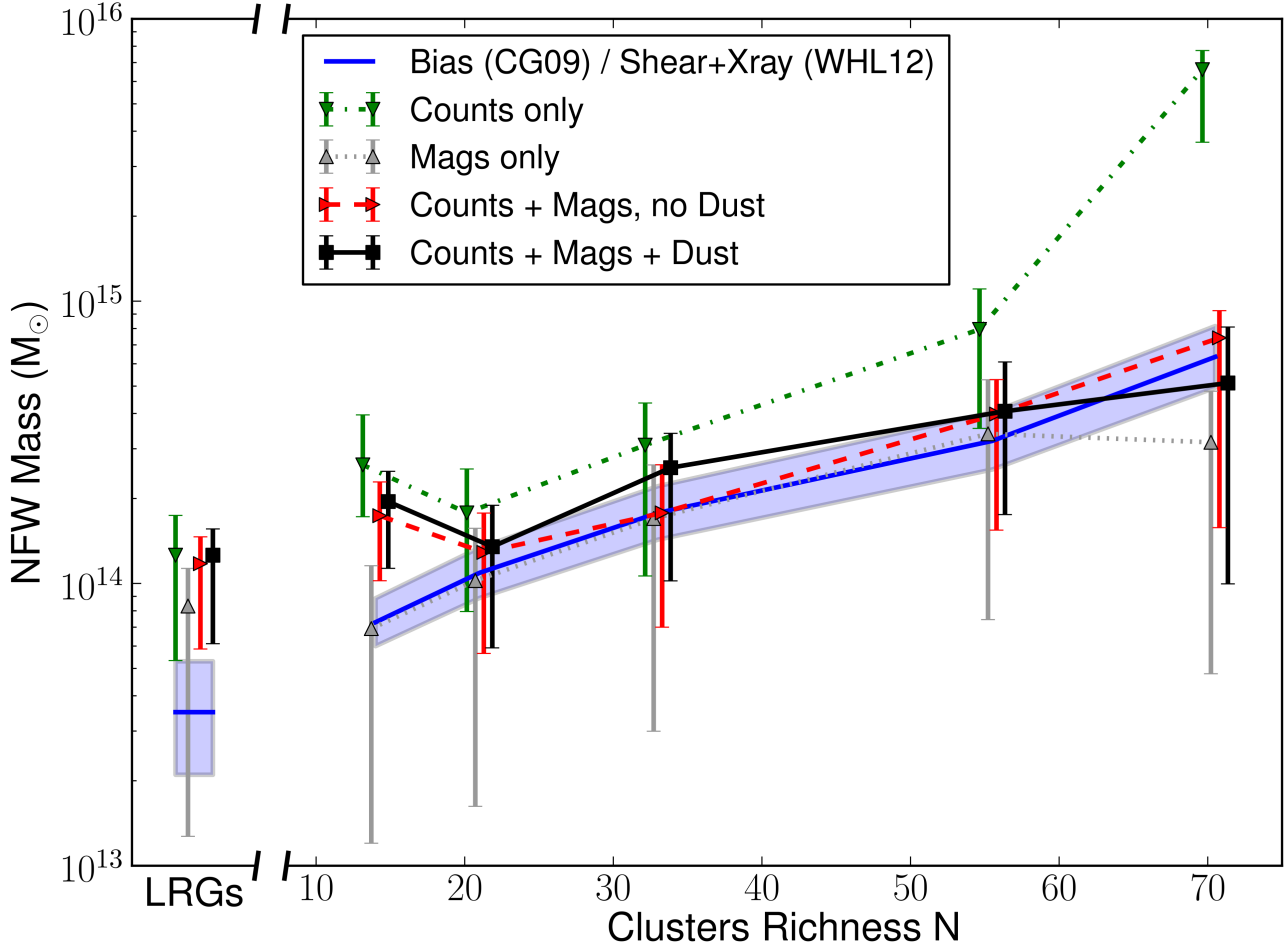}
\caption{Mass-Richness relation for the main NFW halo, compared to the shear and X-ray results given in WHL12 and to the bias results from CG09.  
Different data points represent different combinations of data included in the fit.}
\label{mn_fig}
\end{figure*}

Figure \ref{mn_fig} shows the mass $M_{200}$ of the NFW halo, versus richness $N$.  
The solid blue line shows the masses measured in previous works, with shading indicating the uncertainty.  
The cluster mass-richness relation shown is provided in \cite{dr9sz} (WHL12), determined using weak 
lensing shear and X-ray data.  The values plotted for each richness bin are the mean mass expected given the distribution 
in richness of the clusters in that bin.  
The LRG measurement is based on the measured bias of this sample of 
b=1.85$\pm$0.25, determined using the anisotropic correlation function $\xi(\sigma,\pi)$ in 
\cite{cabre09} (CG09).  This bias can be converted to a halo mass according to \cite{tinker10}, 
resulting in a mass of 
$3.5^{+1.8}_{-1.4}\mbox{\sc{e}}13 \ \mathrm{M}_{\odot}$.  Assuming a finite mass range with upper limit given by the 
smallest cluster halo mass, compared to assuming a narrow mass bin, only changes the mass 
estimate by $1\mbox{\sc{e}}12 \ \mathrm{M}_{\odot}$.

The four sets of data points show the fit results using all data (black squares), $\Sigma_{c}$ and $\Sigma_{m}$ ignoring dust extinction effects 
(red right-facing triangles), only $\Sigma_{m}$ (gray up-facing triangles), and only $\Sigma_{c}$ (green down-facing triangles).  
The results are mainly consistent;  however, fitting only $\Sigma_{c}$ systematically leads to a larger best-fit halo mass.  The $\Sigma_{m}$-only 
results have very large error bars, as expected from the systematic photometry errors in the data.  
The plot points are slightly offset in the x direction for clarity.

\subsection{Dust-to-mass fraction}

Table \ref{results_table} lists the best-fit dust-to-mass ratio $\Gamma$ for the different richness bins.  The measurements are consistent with 
zero, and are therefore upper limits.  The $3 \sigma$ upper limits for the richness bins are 
[2.5, 2.9, 3.7, 1.9, 3.4, 5.5] $\times 10^{-5}$. The best-fit values imply a small dust correction to the mass density profile $\Sigma$;  for 
our most massive lens bin the dust reduces $\Sigma_{c}$ and $\Sigma_{m}$ by less than 5\%.

\subsection{Lens bias}
\label{twohalosection}
We fit the amplitude of the second halo term as parameterized by the bias of the lens.  As the signal to noise 
is very low in the angular range where the second halo term dominates, we do not find interesting constraints 
on the bias.  The different bins' best-fit biases are similar, with 1$\sigma$ uncertainties ranging between 
biases of typically 2 to 9.  A large bias corresponds to a small second halo term;  our best-fit biases 
trend high because the signal at large radii is consistent with zero.  As we do not constrain well 
the bias, we rerun the analysis with a fixed value of $b=3.0$ for the clusters and $b=2.0$ for the LRGs, 
and find almost identical results for the remaining fit parameters.  
These are the results shown in Table \ref{results_table} and Figures \ref{halo_lrgs_fig}-\ref{halo_n5_fig}.

\subsection{Systematic Checks}

\subsubsection{Intrinsic correlations \label{overlap_section}}

The MegaZ LRG sample's redshifts are estimated using photometry, which at times leads to significant redshift error.  Because MegaZ is selected using the 
same criteria as the 2SLAQ spectroscopic sample, the photometric redshift errors in MegaZ can be quantified reliably.  
The true redshift distribution N(z) of the 2SLAQ sample is compared in detail with the photo-z distribution for the MegaZ catalog, as calculated with BPZ, 
in \cite{pol}.  A fit to the distribution, 
consisting of four Gaussian profiles, is shown in Figure \ref{nofz_fig}.  We use this fit to estimate the fraction of the MegaZ sample that lies below redshift 0.4, therefore overlapping the lens sample, taking into account the error on the Gaussian parameters.  We find an overlap rate of 0.012 $\pm$ 0.001.

To assess the level of contamination in the magnification signal due to this overlap, we perform the analysis replacing the MegaZ data with a sample of 
LRGs with spectroscopic redshifts 0.3$<$z$<$0.4.  For this ``overlap" sample we use the same catalog used for the LRG lenses, but limited to this redshift 
range, which as seen in Figure \ref{nofz_fig} is the redshift region with significant overlap between the samples.  
As the MegaZ is an LRG catalog, we expect our LRG spectroscopic galaxies to have similar clustering properties to the MegaZ's redshift outlier LRGs.  
More importantly, this overlap sample strongly correlates spatially with our lenses, as would be expected for any outlier MegaZ galaxies in our 
source sample that lie in this lower redshift range.  As a result, when we perform our stacking analysis using this overlap sample we find a strong 
signal reflecting the intrinsic correlation in counts, magnitudes, and color for galaxies spatially close to the lenses.  This spurious signal rises as 
the galaxies' angular separation decreases, qualitatively similarly to the lensing signal.  
As this spurious signal indicates the strength of the correlation for a completely overlapping redshift sample, we multiply by the overlap fraction of 
0.012 to determine the amount of signal in our stacking analysis that is due to the overlapping fraction of sources.  We note that this scaling by 0.012 
simply scales the total fraction of outlier sources;  the effects of increased contamination as one approaches the lenses is preserved in our estimate 
of the intrinsic correlations.  
The resulting contamination is small:  less than 0.3 times the error bars in $\Sigma_c$, 0.05 times the error bars in $\Sigma_m$, and 
0.02 times the error bars in E(B-V).  We could subtract the intrinsic contribution from the 
magnification data;  however, there are significant error bars on the intrinsic measurements that would add to the analysis' covariance.  Because the 
effect is not significant compared to the current measurement errors we therefore ignore it in the analysis.  In future work with better precision, the 
intrinsic overlap may play an important role.

\subsubsection{NFW concentration relation}

As a check to ensure that our results are not sensitive to our assumptions about the NFW halo concentration, we perform the analysis 
again using the concentration relation from \cite{duffy08}:
\begin{equation}
c(M_{200},z) = 5.71 \left(\frac{M_{200}}{2 \times 10^{12} h^{-1} M_{\odot}}\right)^{-0.084} ( 1 + z )^{-0.47}.
\end{equation} 

Our results are consistent with those using the concentration from \cite{mandelbaum08}.  In most cases the NFW and BCG best-fit masses change by 
less than 10\%, and in all cases well within the $1 \sigma$ errors.  Changing the concentration relation causes the best-fit masses to increase in 
some bins and decrease in others, and does not have a coherent effect on the results.

\subsubsection{Miscentering \label{miscentering_section}}

\begin{figure}
\begin{center}
\includegraphics[width=85mm]{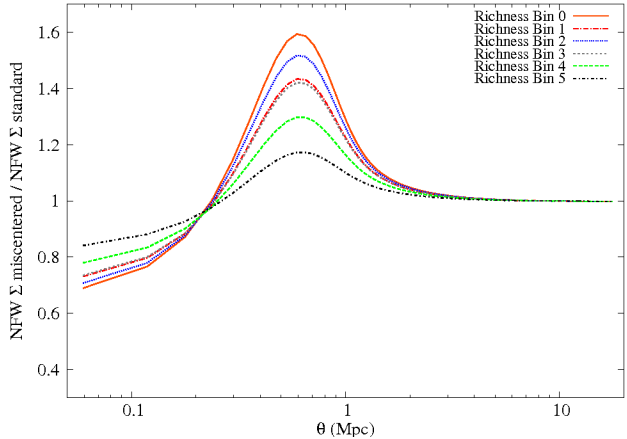}
\end{center}
\caption{Model projected mass density $\Sigma$ of the best-fit NFW profile including miscentering effects, divided by the NFW profile with identical mass without miscentering.  The different lines correspond to the best-fit profiles for different richness bins.}
\label{miscenter_sigma_fig}
\end{figure}

Our implementation of the NFW halo fit includes the effects of halo miscentering as parameterized by \cite{johnston07}, assuming a probability 
of miscentering that depends on the halo richness.  We remove the miscentering 
correction and repeat the analysis to quantify its influence on our results.  
For reference, the effect of miscentering on the model NFW profile is shown in Figure \ref{miscenter_sigma_fig}, in which is plotted the NFW profile 
mass density $\Sigma$ including the contribution from miscentering, divided by the NFW profile $\Sigma$ corresponding to the same halo mass without 
a miscentering contribution.  The different lines represent the best-fit NFW masses for our six richness bins, showing that miscentering tends to 
most strongly affect the least massive haloes.  Miscentering decreases the mass density profile at small radii by several tens of percent, and increases 
$\Sigma$ at intermediate radii by up to $\sim$50\%, causing an overall flattening of the profile shape.  The effect can be quite strong, but 
the decrement at low radii counters the increment at intermediate radii such that the total NFW fits are not significantly altered.  
We find that without miscentering, our best-fit NFW masses change by typically $\sim$10\%, in some bins increasing and in others decreasing.  
The BCG best-fit masses do show a coherent response, decreasing to typically 70-75\% their original values.  
Qualitatively, this behavior is expected because the BCG contribution is significant only on scales where miscentering decreases the NFW profile, 
and therefore removing miscentering will increase the inner NFW $\Sigma$ and lessen the need for a BCG mass contribution.  
Quantitatively, the change is within the BCG $M_{200}$ $1 \sigma$ measurement errors in all bins.

\section{Discussion and Conclusions}

We measure the effects of gravitational lensing magnification on the MegaZ catalog, a photometric sample of LRGs from SDSS DR7.  
We study the effects of gravitational lensing on the 
detected number counts of sources and also the average brightness of sources behind lenses compared to the average values in the source catalog.  
As lenses, we use the WHL12 catalog of galaxy clusters detected in SDSS DR9 \citep{dr9sz}, and also a spectroscopic sample of LRGs selected from 
SDSS DR6.  

We fit the lens NFW halo mass, BCG mass, lens bias, and dust-to-mass ratio in six bins of richness.  
We measure a NFW mass vs. cluster richness relation that is consistent with that 
given in \cite{dr9sz} from weak lensing shear and X-ray data.  Fitting only the counts magnification data, without including flux magnification or 
taking into account reddening measurements, yields systematically larger masses that are in worse agreement with the previous results.

The fit to our lens model is performed between the radial range of 0.04 Mpc to 13 Mpc.  Below roughly 0.04 Mpc we expect the foreground lens galaxy 
(the LRG or the cluster BCG) to obstruct the 
observations of background galaxies, adversely affecting the accuracy of the detection and photometry of the sources \citep{mandelbaum06}.  Such 
data reduction difficulties due to galaxy proximity and crowded fields are not fully understood, and likely bias our results at 
small radii from the lens.  These uncertainties will most strongly affect the best-fit BCG masses;  therefore, we note that we expect that 
the BCG model fits are significantly 
affected by this systematic error.  For example, the best-fit BCG masses using the $\Sigma_{c}$ data alone range 
from 2.2e13 to 4.4e14, indicating a discrepancy in the $\Sigma_{c}$ and $\Sigma_{m}$ data 
\citep[as expected since data systematics affect the different probes 
differently;  see][]{mandelbaum06} and yielding in this case values that are unphysically massive.  It is possible that such systematics also 
contribute to the discrepancy in NFW halo mass when fitting only $\Sigma_{c}$ compared to the full data set.  Our results therefore stress the 
value of multiple probes in analyses of public data where the image reduction can not be fully explored and optimized for each science case.

In addition to measuring the change in number counts and brightness, we measure the change in color of sources behind lenses in order to measure 
reddening of the objects due to dust in the lens systems.  
Dust extinction serves to redden the sources and also make them appear fainter;  the change in flux due to dust is therefore degenerate with the 
change in flux due to lensing.  The degeneracy can be broken through the color dependence of dust extinction, as lensing magnification is achromatic.  

We find no significant reddening of the lensed sources, yielding upper limits on the dust-to-mass ratios of the lenses.  
The systematic errors in the SDSS photometry are large enough to obscure the small color changes expected due to dust extinction.  
To improve on these measurements we will have to be extremely careful about 
quantifying systematic errors in the photometry, as percent-level shifts can arise from (for example) background subtraction errors in crowded fields\footnote[7]{http://www.sdss.org/dr7/products/images/index.html\#skylevz}.  As our cluster lenses themselves constitute crowded fields, the amplitude of 
such photometric errors in source magnitudes can be spatially correlated with the galaxy lens 
positions.

Previous studies imply dust-to-mass ratios for LRGs consistent with our upper limits.  
\cite{barber07} selected LRGs from SDSS DR4 with the same criteria as our LRG lens sample, which was selected 
from SDSS DR6.  Their population synthesis analysis of the sample determined a median extinction in the z band of $A_{z} = 0.0026$;  
this measurement implies E(B-V) $\sim$ 0.001 (much smaller than our systematic color error of 0.02), and 
dust-to-mass ratio $\Gamma$ = 1e-06, assuming our LRG lens best-fit mass.  \cite{tojeiro09} used the VESPA algorithm to determine 
a typical V-band optical depth $\tau_{V} < 0.2$ due to dust in the interstellar medium of SDSS spectroscopic LRGs in SDSS DR7, also selected 
identically to our LRG lens sample.  As optical depth $\tau_{V} = \mathrm{ln}(10)/2.5\ A_{V}$, we can convert this $\tau_{V}$ limit to an E(B-V) 
$<$ 0.06, and to a dust-to-mass ratio $\Gamma <$ 8e-5.  
Using quasar magnification and reddening, \cite{menard09} found a dust-to-mass ratio in the main galaxy sample of SDSS to be roughly 1e-5;  as the 
galaxy sample they studied is expected to be more dusty on average than LRGs, we would expect this to be an upper limit of the dust content in 
our lenses.  We find $3\sigma$ upper limits for the dust-to-mass ratios in the range of 2 to 6 e-5 for the different richness bins, 
in good agreement with expectations for single LRGs.  
We do not find significant evidence for a change in dust-to-mass ratio with lens richness.  
It is expected that cluster environments are less dusty than individual galaxies \citep{chelouche07, muller08};  as our results place only upper 
limits on the dust fraction, we can not support or refute this claim.

Photometric LRGs are a convenient sample to study as they are numerous, bright, and have similar intrinsic colors.  
Their homogenous color is advantageous for two reasons:  
it decreases the noise in the $g-i$ reddening measurements, and it causes LRGs to be well-suited to photometric redshift estimation.  Knowledge 
of the photometric redshift error is essential in studies of gravitational lensing magnification, as the presence of photo-z outliers can mimic the lensing signal.  
The existence of a spectroscopic galaxy sample that is representative of the photometric catalog, as 2SLAQ is for the MegaZ data set, is crucial 
for understanding the true redshift distribution of the photometric galaxies.  Because of these data we can measure very precisely the fraction of overlap 
between our source and lens samples, and therefore constrain the contribution of intrinsic overlap to our lensing signal.  In the current analysis, 
such intrinsic correlation is small compared to the statistical error in the measurements.  It is important, however, that we are able to quantify 
the contribution.

The signal to noise of the lensing-induced counts overdensity is larger than that of the average flux increase of the lens sample.  Although we do measure significant 
magnitude magnification signal, it is typically in the inner angular regions that are dominated by the BCG.  The upper limits on the magnitude 
magnification, however, do yield a constraint on the mass profiles of the lenses.  This effect is most strongly seen in the highest richness bin, where 
the counts alone yield a significantly higher best-fit NFW mass than the combined fit with magnitudes, or with magnitudes and dust.  (All three results 
give a good $\chi^{2}$ fit to the data.)  
Although the signal to noise of the change in counts is larger than that of the change in magnitudes, the factors that convert these basic measurements 
into mass density profiles are better constrained for magnitudes magnification than for counts.  For example, the uncertainty on alpha is six times 
larger for counts due to the higher intrinsic correlation of counts compared to magnitudes.  For the same reason, the covariance between angular bins is 
much higher in the mass profile measured using counts, compared to using magnitudes.  Because of these considerations, relatively poor measurements of 
magnitude magnification can still yield an important constraint on the lens mass profiles.  The contribution of magnitude and reddening measurements 
to the best-fit NFW halo masses, despite the low signal to noise of the measurements, indicates the importance of including these probes in weak lensing 
magnification analyses.

This is the first analysis to simultaneously fit the magnification effects on number counts and magnitudes as well as reddening due to dust 
in order to constrain the mass profiles and dust content of a lens sample.  In addition, this work uses as its source sample a catalog of photometric 
LRGs, which is a bright galaxy population that is numerous and easily-detected in typical survey data.  
We present significant measurements of the mass profiles of our LRG and galaxy cluster lenses that are 
in good agreement with previous estimates using different techniques.  The quality of our measurements is currently limited by the photometric 
error in the source catalog.  Upcoming projects such as the Dark Energy Survey (DES)\footnote[8]{http://www.darkenergysurvey.org}, 
Hyper Suprime-Cam Survey (HSC)\footnote[9]{http://www.naoj.org/Projects/HSC}, and the Large Synoptic Survey Telescope (LSST)\footnote[10]{http://www.lsst.org} will 
produce vast galaxy catalogs with high-quality photometry on which we can apply these analysis techniques to study lensing magnification with unprecedented precision.

\section*{Acknowledgments}

We thank Arnau Pujol for help relating the halo mass with the bias of the LRG lens sample.  
We acknowledge the support of the JAE-Doc (CSIC) fellowship, 
the Spanish Science Ministry
AYA2009-13936, AYA2012-39559, Consolider-Ingenio CSD2007-00060, project
2009SGR1398 from Generalitat de Catalunya, 
the Marie Curie European Reintegration Grant PERG07-GA-2010-268290, 
and the European Commission Marie Curie Initial Training Network CosmoComp (PITN-GA-2009-238356)

\bibliography{mn-jour,bibliography}

\label{lastpage}

\end{document}